\documentclass{aa}
\input{epsf}
\newcommand{\EQ}{\begin{equation}}
\newcommand{\EN}{\end{equation}}
\newcommand{\EQA}{\begin{eqnarray}}
\newcommand{\ENA}{\end{eqnarray}}
\newcommand{\eq}[1]{(\ref{#1})}
\newcommand{\Eq}[1]{Eq.~(\ref{#1})}
\newcommand{\Eqs}[2]{Eqs.~(\ref{#1}) and~(\ref{#2})}

\newcommand{\Fig}[1]{Fig.~\ref{#1}}

\newcommand{\Figs}[2]{Figs~\ref{#1} and \ref{#2}}

\newcommand{\bra}[1]{\langle #1\rangle}

\newcommand{\uu}{\mbox{\boldmath $u$} {}}

\newcommand{\FF}{\mbox{\boldmath $F$} {}}

\newcommand{\grav}{\mbox{\boldmath $g$} {}}
\newcommand{\nab}{\mbox{\boldmath $\nabla$} {}}

\newcommand{\SSSS}{\mbox{\boldmath ${\sf S}$} {}}
\newcommand{\DD}{{\rm D} \, {}}
\newcommand{\dd}{{\rm d} {}}

\newcommand{\ea}{{\rm et al. }}

\def\half{{\textstyle{1\over2}}}

\def\twothird{{\textstyle{2\over3}}}

\newcommand{\yapj}[3]{ #1, {ApJ }{#2}, #3}

\newcommand{\yana}[3]{ #1, {A\&A }{#2}, #3}

\newcommand{\yjfm}[3]{ #1, {JFM }{#2}, #3}

\newcommand{\yprs}[3]{ #1, {Proc. Roy. Soc. Lond. }{#2}, #3}

\newcommand{\ymn}[3]{ #1, {MNRAS }{#2}, #3}

\newcommand{\yjcp}[3]{ #1, {J. Comput. Phys. } {#2}, #3}
\newcommand{\yjour}[4]{ #1, {#2} {#3}, #4}

\newcommand{\yproc}[5]{ #1, in {#3}, ed. #4 (#5), p.#2}

\begin{document}
%\thesaurus{12(02.08.01; 02.20.1; 08.02.1)}
\title{Evolution of highly buoyant thermals in a stratified layer}
\author{Axel Brandenburg\inst{1,2} and John Hazlehurst\inst{3}}
\institute{
NORDITA, Blegdamsvej 17, DK-2100 Copenhagen \O, Denmark
\and Department of Mathematics, University of Newcastle upon Tyne, NE1 7RU, UK
\and Hamburger Sternwarte, Gojenbergsweg 112, D-21029 Hamburg, Germany
}

\date{Received 7 August 2000 / Accepted 7 February 2001}

\abstract{
The buoyant rise of thermals (i.e.\ bubbles of enhanced entropy, but
initially in pressure equilibrium) is investigated numerically in three
dimensions for the case of an adiabatically stratified layer covering
6--9 pressure scale heights. It is found that these bubbles can travel to
large heights before being braked by the excess pressure that builds
up in order to drive the gas sideways in the head of the bubble. Until
this happens, the momentum of the bubble grows as described by the
time-integrated buoyancy force. This validates the simple theory of bubble
dynamics whereby the mass entrainment of the bubble provides an effective
braking force well before the bubble stops ascending. This is quantified
by an entrainment parameter alpha which is calculated from the simulations
and is found to be in good agreement with the experimental measurements. This
work is discussed in the context of contact binaries whose secondaries
could be subject to dissipative heating in the outermost layers.
}

\maketitle

\section{Introduction}

Highly buoyant bubbles with large specific entropy excess relative to
the surroundings have been invoked by Hazlehurst (1985) in an attempt
to explain the almost equal effective temperatures of the two components
of contact binaries.

As noted by Sinjab \ea (1990) there are strong parallels between the
highly buoyant bubbles of Hazlehurst and the `thermals' found to occur
in the earth's atmosphere. Subsequently, Hazlehurst (1990) confirmed the
existence of a formal relationship between Turner's (1963) treatment
of thermals, involving entrainment of matter, and his own treatment of
highly buoyant bubbles (`interlopers') in contact binaries, as these
bubbles annex new material.

In this paper we shall discuss the question of the validity, from a
fluid-dynamical standpoint, of the simple bubble or thermal picture. We
shall then go on to show how it is possible to determine numerically the
value of the entrainment coefficient $\alpha$ (here called
$\alpha_{\rm v}$) which enters Turner's and several other investigations;
the determination of a related coefficient (here called $\alpha_{\rm m}$)
entering the Hazlehurst theory is also discussed.

We believe this to be the first attempt to evaluate the (previously
semi-empirical) entrainment coefficient $\alpha$ on a fluid-dynamical
basis.

\section{Model setup}

We adopt a basic setup of our model that is similar to that used
normally to study convection in a stratified plane-parallel layer between
impenetrable boundaries (e.g.\ Hurlburt \ea 1984). In particular, we use
stress-free boundary conditions at the top and bottom, with a prescribed
flux $F$ at the bottom and a prescribed temperature $T_{\rm top}$ at the
top. Here, however, we assume the thermal equilibrium stratification to be
adiabatic, so it is marginally stable to the onset of convection. Thus,
when we insert a hot (buoyant) bubble, it will rise unaffected by the
stratification (except for effects related to the growth and expansion
of the ascending bubble), so there is no restoring force acting on the
bubble as it rises.

\subsection{Adiabatic stratification}

Hydrostatic equilibrium requires that the weight of the atmosphere is
balanced by the pressure gradient. However, if the entropy is constant, the
pressure gradient can be written as $\rho^{-1}\nab p=\nab h$, where $p$
is pressure, $\rho$ is density, and $h$ is enthalpy. We adopt a perfect
gas for which $h=c_{\rm p}T$, where $c_{\rm p}$ is the specific heat
at constant pressure. For constant gravitational acceleration $g>0$,
this implies a {\it linear} temperature stratification that is given by
\EQ
{\dd\over\dd z}(c_{\rm p}T)=g,
\EN
where $z$ is depth, which is assumed to increase {\it downwards}. Thus,
the vertical temperature profile of the basic state is given by
\EQ
T=T_{\rm top}+(z-z_{\rm top})g/c_{\rm p}.
\label{Tstrat}
\EN
In the absence of any motion there is only radiative flux,
$\FF$, for which we adopt the diffusion approximation, so
\EQ
\FF=-K\nab T,
\label{Frad}
\EN
where $K$ is the radiative conductivity. Thermal equilibrium requires
$\nab\cdot\FF=0$, so the $z$-component of the flux is constant but,
because the temperature gradient is constant, this is only possible
if $K={\rm const}$. We now need to discuss the choice of $K$ and other
parameters.

\subsection{Choice of parameters}

We adopt nondimensional units by defining a unit length $d$, and our
bubble will usually have the initial radius $R_0=d$ (although we present
initially some cases where $R_0=0.5\,d$. We measure time
in units of $(d/g)^{1/2}$, density in units of $\rho_0$ (we choose
$\rho=\rho_0$ at the location where the centre of the bubble will
be introduced), and specific entropy, $s$,
in units of $c_{\rm p}$. This corresponds to setting
\EQ
d=g=\rho_0=c_{\rm p}=1.
\EN
A relevant nondimensional quantity is the normalised input flux
\EQ
{F\over\rho_0(gd)^{1/2}}\quad\mbox{(nondimensional input flux)}.
\EN
For secondaries of contact binaries this ratio is around
$10^{-3}$. Specifying $F$ fixes $K=c_{\rm p}F/g$. There is however the
numerical constraint that the mesh Peclet number, based on the sound
speed $c_{\rm s}$ and the mesh width $\Delta x$,
\EQ
\mbox{Pe}_{\rm grid}=\Delta x\,c_{\rm s}/\chi,
\EN
should not exceed a certain empirical upper limit of 10--100. This
limit arises from the fact that the advection of sharp structures
tends to generate small ripples on the scale of the mesh which need to be
damped. Here, $c_{\rm s}^2=\gamma p/\rho$, $\gamma=c_{\rm p}/c_{\rm
v}$ is the ratio of specific heats, and $\chi=K/\rho c_{\rm p}$ is the
radiative diffusivity. In the present paper we shall consider values of
$F$ in the range 0.001--0.005. In all cases we assume $\gamma=5/3$.

We adopt cartesian coordinates $(x,y,z)$ where $z$ points downwards. The
bubble centre is placed initially at $x=y=z=0$. In order that the bubble
be initially sufficiently far away from the bottom boundary and that
it can rise over a distance of at least a few times its own radius
(which is unity in most of the models), we chose the vertical extent of
the computational box to be from $z_{\rm top}=-6$ to $z_{\rm bot}=2$.

Next, we fix the temperature stratification within the box by specifying
the value of $T_{\rm top}$, which is held constant by the boundary
condition chosen. The choice of this parameter is restricted by numerical
considerations. Since temperature is proportional to the pressure scale
height, small values of $T_{\rm top}$ imply a short pressure scale height
at the surface. However, numerical stability and accuracy considerations
require that the scale height cannot be less than at least 2--3 mesh
zones. The non-dimensional pressure scale height at the top is
\EQ
\xi_0=H_{\rm p,\,top}/d=(c_{\rm p}-c_{\rm v})T_{\rm top}/gd,
\label{xidef}
\EN
where we have included $d$ and $g$ factors, even though they are unity. For
runs with $N_z=50$ meshpoints in the vertical direction, we were able to
use $\xi_0=0.3$. For orientation, we note that this yields a $\ln(p_{\rm
bot}/p_{\rm top})$ of 6.2 pressure scale heights between top and bottom
of the box. The local pressure scale height at $z=0$ is then given by
$0.4\times|z_{\rm top}|+\xi_0=2.7$. Using Eqs \eq{Tstrat} and \eq{xidef}
we also see that the top of the adiabatic atmosphere would be at
\EQ
z_\infty=z_{\rm top}-2.5\xi_0\quad\mbox{(for $\gamma=5/3$)},
\label{zinfty}
\EN
but this should be outside the computational box. For $z_{\rm top}=-6$
and $\xi_0=0.3$ this gives $z_\infty=-6.75$.

For a perfect gas we have $p/\rho=(c_{\rm p}-c_{\rm v})T$ and
$s=c_{\rm v}\ln p-c_{\rm p}\ln\rho$ (in dimensional form). The density
stratification follows then from hydrostatic balance and the assumption
that the entropy of the unperturbed model is constant:
\EQ
\rho/\rho_0=(1-z/z_\infty)^{1/(\gamma-1)}.
\EN
We mentioned already that we express density in units of $\rho_0$,
which is the density at $z=0$, which is where the bubble will be
positioned. Using this together with the definition of $s$ we can
use \Eq{zinfty} to express the value of the background entropy,
\EQ
s_0=0.6\ln(-0.4z_\infty)\quad\mbox{(for $\gamma=5/3$)}.
\EN
For $z_\infty=-6.75$ this gives $s_0=0.596$.

This completes the definition of the background state. We now turn
to the discussion of the model equations and initial and boundary
conditions adopted.

\subsection{Governing equations}

In the dynamical case the specific entropy is not only affected by the
radiative flux divergence, but also by the rate of viscous dissipation, so
\EQ
\rho T{\DD s\over\DD t}=\nab\cdot K\nab T+2\nu\rho\SSSS^2,
\label{DsDt}
\EN
where ${\rm D}/{\rm D}t=\partial/\partial t+\uu\cdot\nab$ is the lagrangian
derivative, $\nu=\mbox{const}$ is the kinematic viscosity and
\EQ
{\sf S}_{ij}=\half(\partial_j u_i+\partial_i u_j
-\twothird\delta_{ij}\partial_k u_k)
\EN
is the (traceless) rate of strain tensor. Equation \eq{DsDt} is solved
together with the momentum equation,
\EQ
{\DD\uu\over\DD t}=-{1\over\rho}\nab p+\grav
+{1\over\rho}\nab\cdot(2\nu\rho\SSSS),
\label{DuDt}
\EN
and the continuity equation
\EQ
{\DD\ln\rho\over\DD t}=-\nab\cdot\uu.
\label{DlnrhoDt}
\EN

We solve Eqs \eq{DsDt}, \eq{DuDt} and \eq{DlnrhoDt} using the sixth order
compact derivative scheme of Lele (1992) and a third order Hyman scheme
for the time step. For earlier applications of this code see Nordlund \&
Stein (1990) and Brandenburg \ea (1996).

The value of $\nu$ is dictated again by numerical considerations, and
in practice we take $\nu\approx\chi$, but note that $\nu$ is independent
of $z$ whilst $\chi$ is not. In all cases considered below we have used
$\nu=6\times10^{-3}$.

\subsection{Initial and boundary conditions}

The bubble centre is placed initially at $x=y=z=0$ and has an entropy
profile of the form
\EQ
s=s_0+\Delta s\,\Phi(r),
\EN
where
\EQA
\Phi(r)=\left\{
\begin{array}{ll}
\exp\left(-\displaystyle{{r^2\over R_0^2-r^2}}\right)&\quad\mbox{if $r<R_0$}\\
0&\quad\mbox{otherwise}
\end{array}
\right.
\ENA
is a profile function defining the initial shape of the bubble. Here $r$
is the initial distance from the centre. Initial pressure equilibrium
requires that the increase of $s$ is compensated by a corresponding
decrease of $\ln\rho$, so
\EQ
\ln\rho={1\over\gamma-1}\ln\left(1-{z\over z_\infty}\right)
-\Delta s\,\Phi(r).
\EN

The initial entropy excess of the bubble is given by the parameter
$\Delta s$. [In all cases presented we take $\Delta s=0.5$, which
corresponds to the value used by Hazlehurst (1985), who adopts units
where the nondimensional specific entropy is larger by a factor of
5. We note that larger values of $\Delta s$ make the bubble rise faster,
but we found that even for $\Delta s=2$ the motion remained subsonic.]

We assume stress-free boundaries at the top and bottom, and prescribe
the value of $T$ at the top and $\partial T/\partial z$ at the bottom,
as is usual for convection calculations (e.g.\ Hurlburt et al.\ 1984).
We use periodic boundary conditions in the $x$ and $y$ directions. The
horizontal extent of the box, $|x|<L_x$ and $|y|<L_y$, is varied between
$L_x=L_y=4$ and 16.

\subsection{Allowing for three-dimensional effects}

\begin{figure}[h]\epsfbox{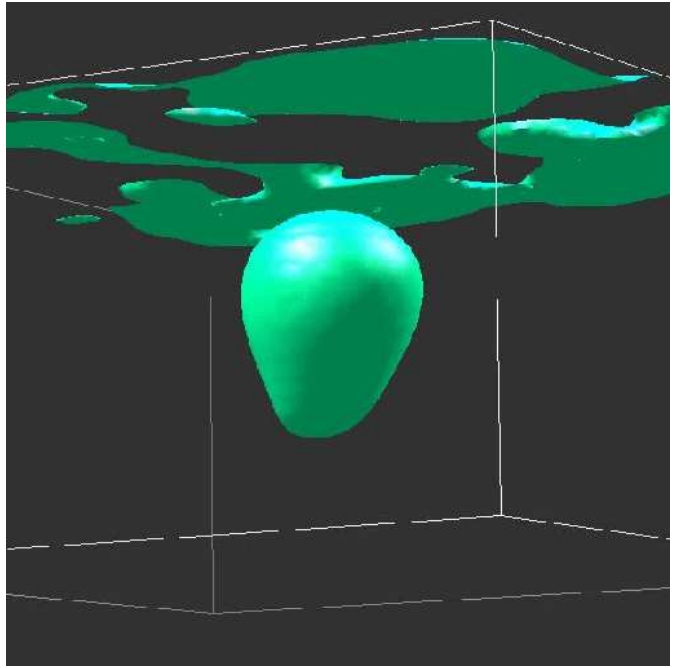}\caption[]{
Three-dimensional representation of the specific entropy.
The initial velocity perturbations are noticeable mostly near the
top layers, but the bubble itself remains fairly axisymmetric.
$100\times100\times50$ meshpoints.
}\label{Fn100x100x50r1}\end{figure}

In order to assess the fragility of the bubble during its ascent we have
adopted in many cases substantial initial velocity perturbations. In
\Fig{Fn100x100x50r1} we show a three-dimensional representation of the
entropy for a run with $\Delta s=0.5$ using initial velocity perturbations
with $u_{\max}/c_{\rm s}=0.4$ and $u_{\rm rms}/c_{\rm s}=0.02$. As is
evident from \Fig{Fn100x100x50r1}, the entropy of the blob is hardly
affected by these perturbations and only near the surface does one see
strong perturbations.

\epsfxsize=8.8cm

\begin{figure}[h]\epsfbox{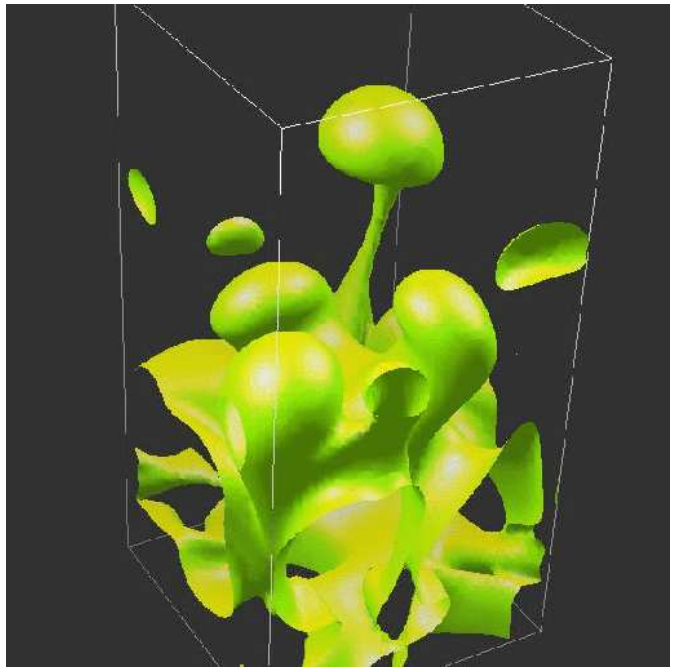}\caption[]{
As \Fig{Fn100x100x50r1}, but for the case of a horizontal layer of
enhanced specific entropy initially. $50\times50\times100$ meshpoints.
}\label{FVAR100_lev0.907}\end{figure}

It is interesting that the bubble remains an entity during much of its
ascent. In fact, even when the initial condition is quite different,
bubble-like structures tend to develop. As an example we show in
\Fig{FVAR100_lev0.907} a case where we have introduced an almost uniform
horizontal layer of enhanced specific entropy with a gaussian vertical
profile initially. We have superimposed random small scale perturbations
to get the buoyancy instability started.

\epsfxsize=8.8cm

In the following we study in more detail the dynamics of an isolated buoyant
bubble or thermal, as it shall also be referred to.

\section{Dynamics of isolated thermals}

In the following we consider cases with different degrees of stratification
and different extents of the computational domain.

\subsection{Modest stratification}

\begin{figure}[h]\epsfbox{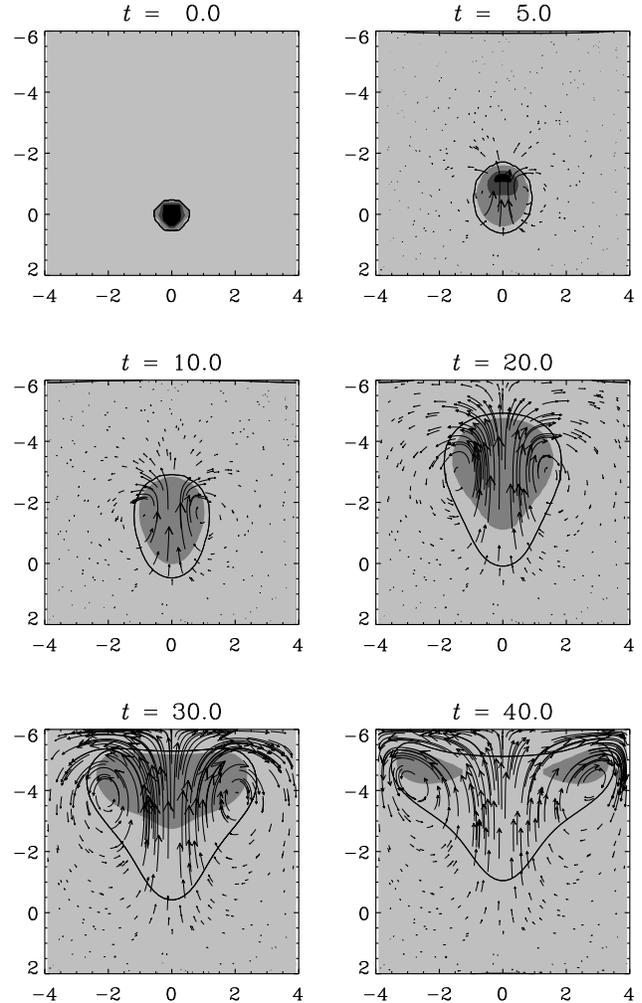}\caption[]{
Velocity vectors superimposed on a grey scale representation of the
entropy (dark indicates high entropy; all panels have the same grey
scale). The velocity is shown in a fixed
frame of reference. The initial specific entropy excess of the bubble
is $\Delta s=0.5$ and its initial radius is $R_0=0.5$. The single contour
shows the position where $s=s_{\rm crit}=0.001+s_0$. $F=0.005$,
$50^3$ meshpoints.
}\label{Fpslice0_n50r1}\end{figure}

We begin by considering first vertical cross-sections of entropy and
velocity; see \Fig{Fpslice0_n50r1}. The archimedian buoyancy force is
largest in the middle of the bubble, and that is also where the vertical
velocity is largest. On both sides of the bubble the velocity turns
over, as expected (compare with observations of thermals described by
Scorer 1957).

\epsfxsize=8.8cm

\begin{figure}[h]\epsfbox{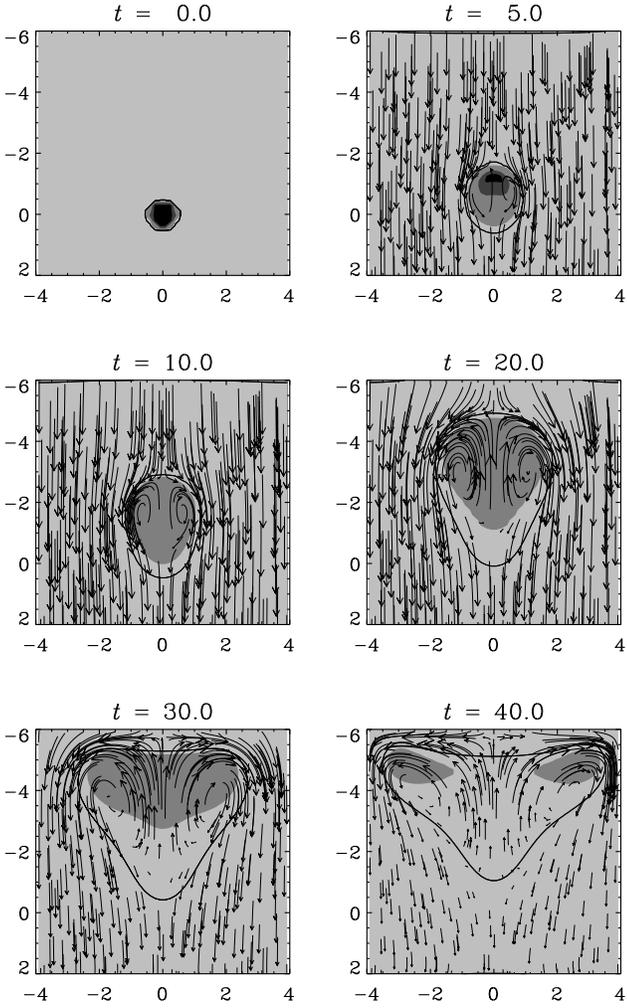}\caption[]{
Same as \Fig{Fpslice0_n50r1}, but now the velocity vectors are shown in
a comoving frame of reference.
}\label{Fpslice0_n50r1_comov}\end{figure}

A rather different impression is obtained when looking at the bubble
in a comoving frame of reference; see \Fig{Fpslice0_n50r1_comov}. In
this frame there is a stagnation point and hence there is a clear
distinction of regions inside and outside the bubble. The flow pattern
generally conforms with the notion of bubbles behaving like balloons
with a more-or-less well defined surface and gas flowing around this
surface. However, the bubble clearly grows in size and even its mass
grows during its ascent.

It should be noted that this addition of new material with `different'
specific entropy does not contradict the bubble concept. This is because
there is no entropy discontinuity between the bubble and the surroundings,
so that the entropy of a captured particle can be changed gradually --
by friction and by thermal diffusion -- as it enters the bubble.

\epsfxsize=8.8cm

\epsfxsize=8.8cm\begin{figure}[h]\epsfbox{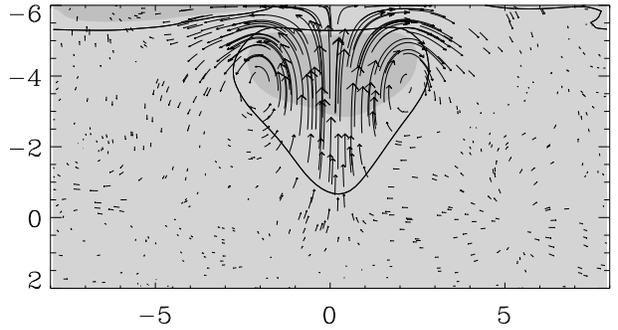}\caption[]{
Same as \Fig{Fpslice0_n50r1}, but for $t=30$ and for a wider box ($L_x=8$).
Although in this calculation strong three-dimensional perturbations have
been applied initially, the effects on the shape of the bubble are negligible.
$F=0.005$, $100\times100\times50$ meshpoints, $\Delta s=0.5$, $R_0=0.5$.
}\label{Fpslice0_n100x100x50r1}\end{figure}

By the time $t=30$ the effects of the lateral boundaries have
begun to affect the evolution of the bubble. Therefore we show in
\Fig{Fpslice0_n100x100x50r1} the case of a wider box ($L_x=8$).
Note that there is now a noticeable flow speed even beyond $|x|=4$
(the extent of the box in the previous case). In this calculation we
have also included strong velocity perturbations, but the overall flow
pattern is still dominated by the rising bubble.

\begin{figure}[h]\epsfbox{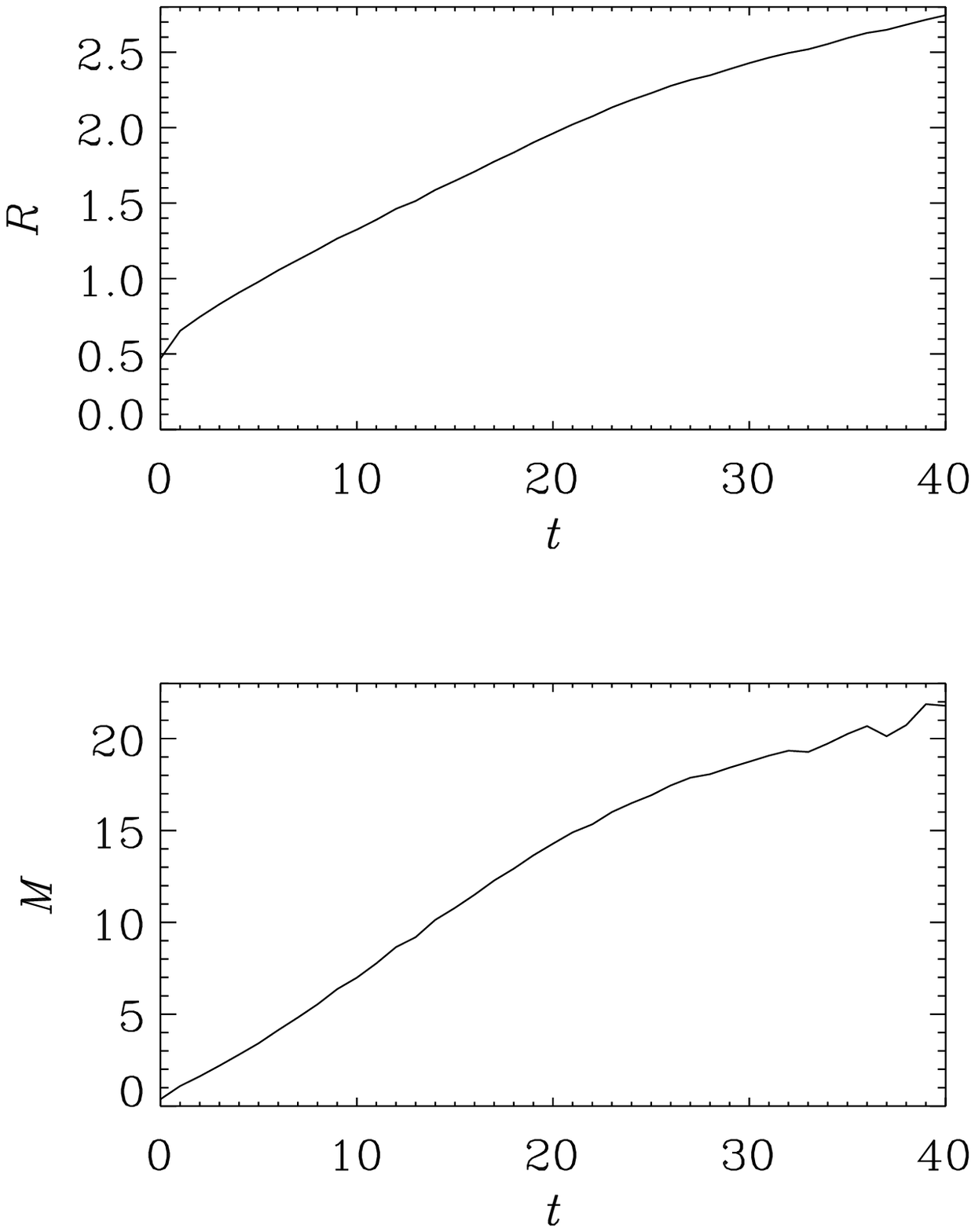}\caption[]{
Radius and mass of the bubble shown in \Fig{Fpslice0_n50r1}.
$F=0.005$, $50^3$ meshpoints, $\Delta s=0.5$, $R_0=0.5$. 
}\label{Fpradius_n50r1}\end{figure}

In order to quantify the rise and the growth of the bubble in detail, we
define the bubble $B$ as all points in space where $s\ge s_{\rm crit}$,
with $s_{\rm crit}$ just a little larger than the background value,
which is here $s_0=0.596$, so we chose $s_{\rm crit}=0.597$. The volume
of the bubble is then estimated as
\EQ
V(t)=\int_BdV,
\EN
and the volume radius of the bubble is
\EQ
R(t)=\left(V/{\textstyle{4\over3}}\pi\right)^{1/3}.
\EN
The mass of the bubble is
\EQ
M(t)=\int_B\rho\,dV.
\EN
In \Fig{Fpradius_n50r1} we plot $R(t)$ and $M(t)$. Both functions increase
monotonically, except for some minor departures at late times when the
bubble has reached the top of the layer. Note also that at early times
($t<4$) $R$ increases somewhat faster than at later times. Qualitatively
this type of behaviour is expected, because radiative diffusion causes
structures to grow proportional to $t^{1/2}$, which causes an infinite
slope of $R(t)$ at $t=0$.

\epsfxsize=8.8cm

\begin{figure}[h]\epsfbox{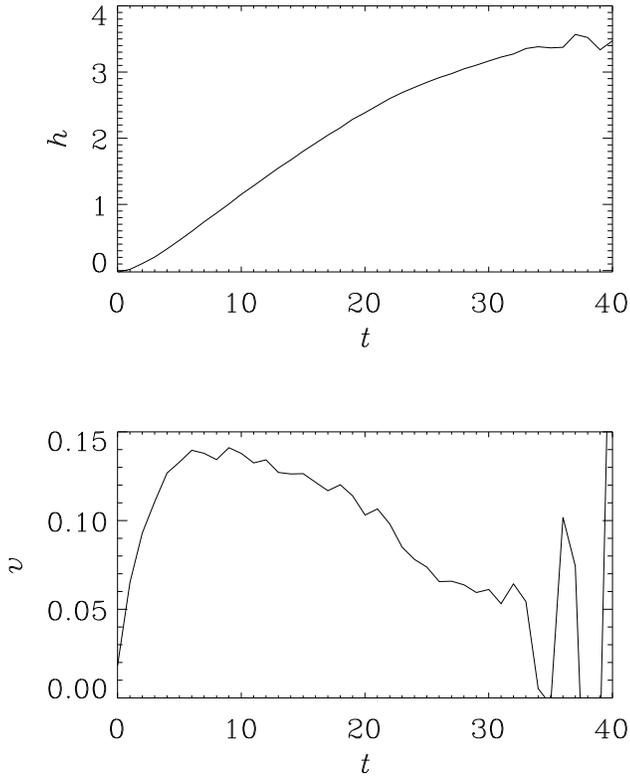}\caption[]{
Height and speed of the bubble shown in \Fig{Fpslice0_n50r1}.
$F=0.005$, $50^3$ meshpoints, $\Delta s=0.5$, $R_0=0.5$. 
}\label{Fpheight_n50r1}\end{figure}

In order to make further comparison with Hazlehurst's theory of buoyant
bubbles, we measure the position of the centre of mass of the bubble,
\EQ
z_{\rm bubble}=\left.\int_Bz\rho\,dV\right/\int_B\rho\,dV.
\EN
Since $z$ decreases upwards we define the height of the bubble as
$h=-z_{\rm bubble}$. The height $h$ and velocity $v=\dd h/\dd t$ of
the bubble are shown in \Fig{Fpheight_n50r1}. Note that the height seems
to approach a maximum near $h=3.5$, so the centre of mass of the bubble
does note quite reach the top of the box. (Below we shall show that for
larger bubbles, $R_0=1$, the maximum height is even less, suggesting that
this is at least partly a geometrical effect; we shall also see that the
top of the bubble does reach the top of the box in all cases.)

\epsfxsize=8.8cm

\epsfxsize=8.8cm\begin{figure}[h]\epsfbox{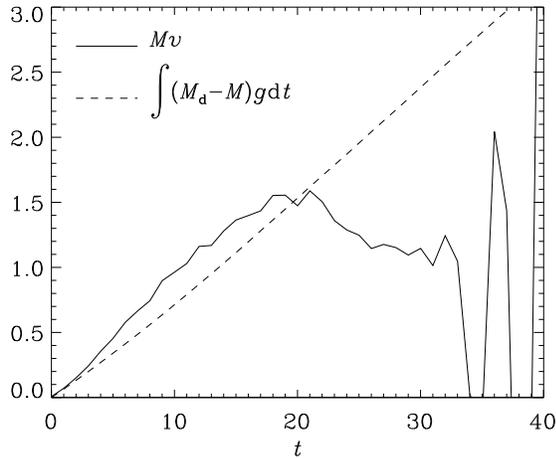}\caption[]{
Momentum and integrated buoyancy force acting on the bubble
shown in \Fig{Fpslice0_n50r1}.
$F=0.005$, $50^3$ meshpoints, $\Delta s=0.5$, $R_0=0.5$. 
}\label{Fpmass_int}\end{figure}

The momentum of the bubble, $Mv$, is plotted in \Fig{Fpmass_int} and
compared with the time integrated buoyancy force,
\EQ
\int_0^t (M_{\rm d}-M)g\,\dd t,
\label{buoyancy_force}
\EN
where
\EQ
M_{\rm d}(t)=\int_B\tilde\rho\,dV
\EN
is the mass of displaced material and $\tilde\rho$ is the density of the
undisturbed medium. Between $t=5$ and 15 the momentum of the bubble is
somewhat larger than expected from the buoyancy force. This discrepancy
depends somewhat on the definition of the boundary of the bubble. However, more
dramatic is the sudden loss of momentum of the bubble after $t\approx20$,
whilst the buoyancy force, as estimated by \Eq{buoyancy_force}, continues
to operate beyond this time.

There are at least two possible reasons for this sudden braking
effect. One reason could be that the blob gets too close to the top
and loses momentum simply because of pressure build-up between the bubble
and the top boundary. A second possibility could be that the bubble
loses momentum due to some genuine resistance mechanism, such as wave braking or
viscous friction. However, the braking effect seen in the simulations
is too sudden and too strong to be explained by any genuine braking
mechanism. Thus, we now turn to the first possibility, of which we can
distinguish two variants. It is possible that the pressure build-up near
the top is either an artifact of the top boundary being impenetrable,
or it could be simply a feature of strong density stratification
which causes the bubble to expand rapidly sideways. In order to drive
strong sideways motions there must naturally be a horizontal pressure
gradient which would also act in the vertical direction and slow down
the ascent. This mechanism is known in compressible convection as buoyancy
braking (Hurlburt \ea 1984).

In order to clarify the nature of the additional braking effect seen
in the simulations, we first compare with a simulation using a somewhat
taller box to see whether or not the braking sets in later, as would be
the case if the impenetrable top boundary was the reason for the braking
effect. (In the following we use calculations where $R_0=1$.)

\subsection{Moving the top boundary further away}

With $\xi_0=0.3$ and $z_{\rm top}=-6.0$ we have $z_\infty=-6.75$, see
\Eq{zinfty}, so we can move the top boundary upwards by no more than
about 10\%.  In the following, we discuss a model with $z_{\rm top}=-6.5$,
but otherwise the same stratification. This means that at the new boundary
we have to change $\xi_0$ by $\Delta\xi=0.4\times\Delta z=0.2$, so we
have to require $\xi_0=0.1$.

\epsfxsize=8.8cm\begin{figure}[h]\epsfbox{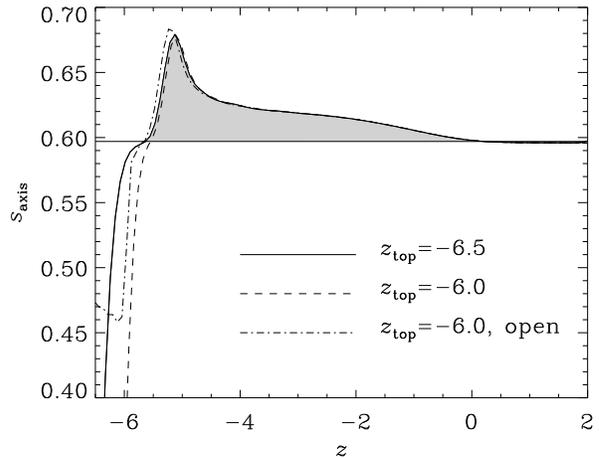}\caption[]{
Vertical profile of $s$ along the axis of the bubble for $t=11$ and
$z_{\rm top}=-6.5$ and $\xi_0=0.1$ (solid line) and $z_{\rm top}=-6.0$
and $\xi_0=0.3$ (dashed and dash-dotted lines for closed and open
boundaries). Note that the entropy profile at the
location of the bubble is not significantly affected by the value of
$z_{\rm top}$. The region where $s>s_{\rm crit}$ is shown in grey. The
entropy drop near the surface is a consequence of fixing the top
temperature, but the location of this entropy drop moves further away
as we extend the box. $F=0.001$, $50^2\times100$ meshpoints, $R_0=1$.
}\label{Fpss_comp}\end{figure}

\epsfxsize=18cm\begin{figure*}[t!]\epsfbox{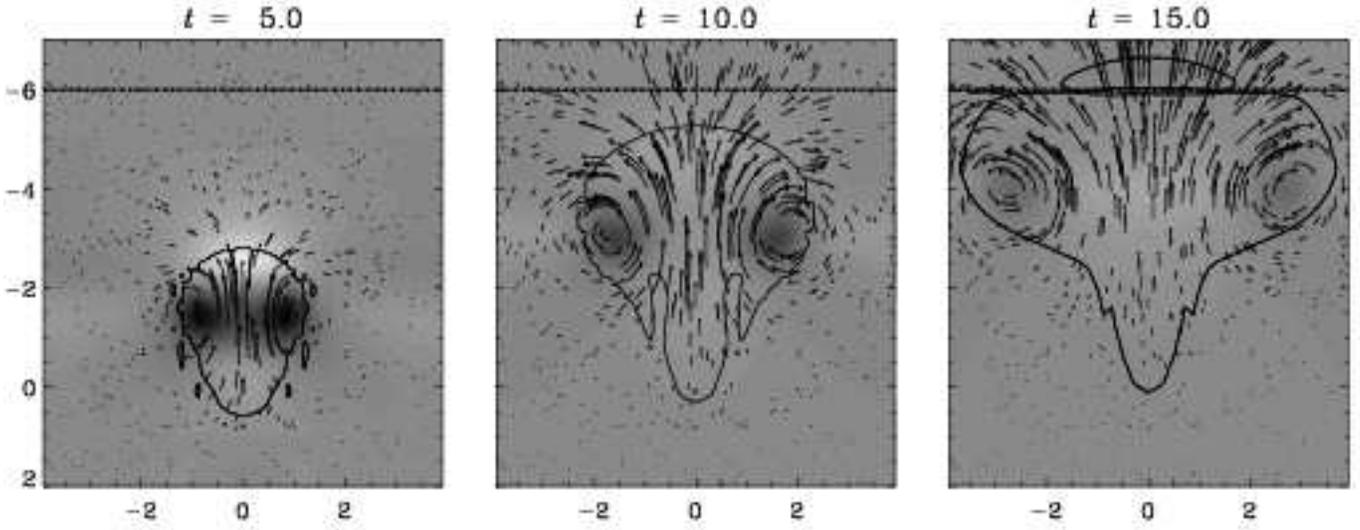}\caption[]{
Grey scale representation of the pressure fluctuation together with
velocity vectors and a contour marking the location where $s=s_{\rm
crit}$. Light refers to high pressure fluctuation and dark to low pressure
fluctuation. The surface at $z=-6$ is marked by a dash-dotted line. Above
this line there is a zero-gravity `buffer layer', modelling the effects
of an open boundary.
}\label{Fppp}\end{figure*}

In \Fig{Fpss_comp} we compare the vertical entropy profiles along the
axis of the bubble for $z_{\rm top}=-6.5$ (with $\xi_0=0.1$) and $z_{\rm
top}=-6.0$ (with $\xi_0=0.3$). We also compare with the case of an open
top boundary that we have modelled by putting an extra layer on top of
the box where gravity goes smoothly to zero and, as in Brandenburg
\ea (1996), radiative diffusion is replaced by a heating/cooling term
of the form $-\tau^{-1}(z)\rho(T-T_{\rm top})$, where $\tau^{-1}=0$
everywhere except above $z=z_{\rm top}$ where it goes smoothly to
$\tau^{-1}=10$. This procedure allows the flow to penetrate the layer
$z=z_{\rm top}$ freely.

It turns out that the entropy profiles at the location of the bubble
are not significantly affected by the properties of the
top boundary. The entropy drop
near the surface is a consequence of fixing the top temperature, $T_{\rm
top}$. Any increase in the logarithmic pressure at the top, $\delta\ln
p_{\rm top}$, causes a corresponding decrease in the entropy, $\delta
s_{\rm top}=-0.4\delta\ln p_{\rm top}$. Note, however, that the location
of this entropy drop at the surface moves further away from the location
of the bubble as we extend the box.

\begin{figure}[h]\epsfbox{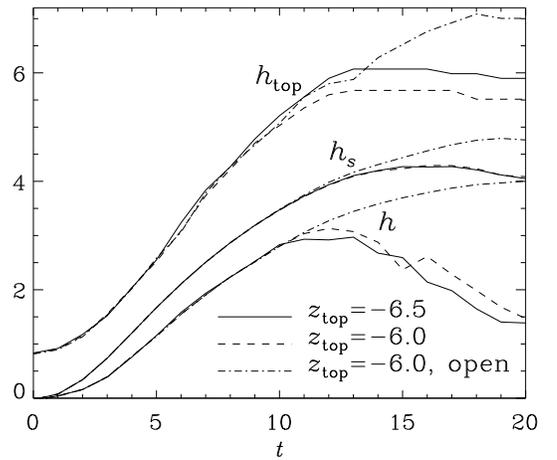}\caption[]{
Height of the bubble (as measured by $h$, $h_s$ and $h_{\rm top}$)
for different values of $z_{\rm top}$. Note that the top of the
bubble, $h_{\rm top}$, reaches the top boundary. $F=0.001$, $50^3$
and $50^2\times100$ meshpoints, $\Delta s=0.5$, $R_0=1$.
}\label{Fph_comp}\end{figure}

In \Fig{Fppp} we show the pressure fluctuations (relative to the
horizontal mean) together with velocity vectors. Near the top of the
bubble there is a strong local maximum of the pressure fluctuation
that drives the gas sideways. We have checked that the ram pressure
integrated over the projected surface is roughly what is needed to
explain the discrepancy between acceleration and buoyancy force. This
is suggestive of buoyancy braking being the cause of the sudden drop of
momentum seen in \Fig{Fpmass_int} (for $R_0=0.5$) and in \Fig{Fph_comp}
(for the present case of $R_0=1$).

In \Fig{Fph_comp} we compare the evolution of $h$ in the two cases
with different values of $z_{\rm top}$. Within
the range of accuracy the two curves are consistent. Of course, the
difference in the value of $z_{\rm top}$ is not very large, but the
increase in the total number of scale heights covered in the simulation
is significant: the value of $\ln(p_{\rm bot}/p_{\rm top})$ has increased
from 6.2 to 9.0 pressure scale heights.

\epsfxsize=8.8cm

Note that $h$ reaches a maximum around $3$, i.e.\ significantly less
than the value of $|z_{\rm top}|$. This is partly a geometrical effect,
because smaller bubbles are able to travel somewhat further (in the
previous subsection, where $R_0=0.5$ instead of 1.0, the bubble went
to $h=3.5$). In the case of the open top boundary, the bubble rises
till $h=4$, which is still small compared with $|z_{\rm top}|=6$.
The relatively small values of $\max(h)$ are partly due to the fact
that $h$ measures the position of the centre of mass, but in the
strongly stratified case most of the mass is at the bottom of the
bubble. Furthermore, in all cases considered the bubbles take mushroom
form with a significant portion of the mass residing in the stem of
the mushroom. A somewhat better representation of where most of the
hot material resides is gained by looking at the value of the entropy
weighted height, $h_s=-z_s$, where
\EQ
z_s=\left.\int_Bz(s-s_{\rm crit})\rho\,dV\right/
\int_B(s-s_{\rm crit})\rho\,dV,
\EN
which is also plotted in \Fig{Fph_comp}. However, even the entropy
weighted height of the bubble is not very close to the top of the
bubble. Nevertheless, the location $h_{\rm top}$ of the top of the bubble,
i.e.\ where $s=s_{\rm crit}$, does reach value close to $|z_{\rm top}|$;
see \Fig{Fph_comp}. There remains however some worry that the centre of
mass of the bubble is generally unable to travel great distances. In
order to clarify this possibility we now consider the case of weak
stratification where we can easily increase the extent of the box.

\subsection{Weak stratification}

We now consider a case of weak stratification and two different values of
$z_{\rm top}$ ($-6$ and $-14$); hence we choose $\xi_0=30$ and $\xi_0=26.8$
respectively, so that the background stratification is the same for
$-6<z<2$. The total number of pressure scale heights in these two cases
is $\ln(p_{\rm bot}/p_{\rm top})=0.25$ and 0.5, respectively.

\epsfxsize=8.8cm\begin{figure}[t!]\epsfbox{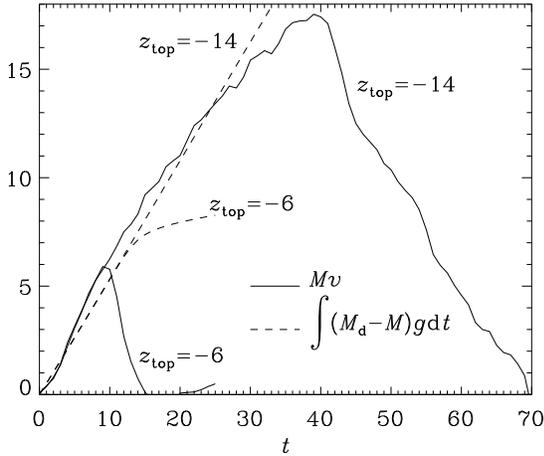}\caption[]{
The effect of the vertical extent of a weakly stratified box on the balance between
momentum and integrated buoyancy force acting on the bubble.
$F=0.005$, $50^3$ and $50^2\times100$ meshpoints, respectively.
$\Delta s=0.5$, $R_0=1$.
}\label{Fpmass_int_comp}\end{figure}

\begin{figure*}[t!]\epsfbox{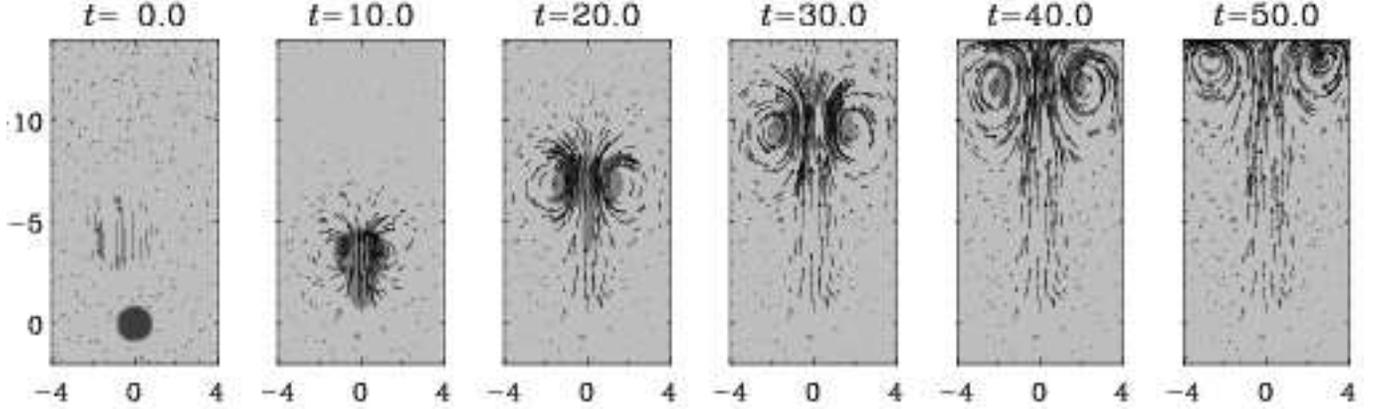}\caption[]{
Velocity vectors superimposed on a grey scale representation of the
entropy for a tall box with weak stratification.  The velocity is
shown in a fixed frame of reference. (The velocity vectors at $t=0$ at
$z\approx-4$ result from the initial perturbation.) The initial specific
entropy excess of the bubble is $\Delta s=0.5$ and its initial radius
is $R_0=1.0$. $50^2\times100$ meshpoints.
}\label{Fpslice0_n50r1_rad1part_xi30_longer}\end{figure*}

\begin{figure*}[t!]\epsfbox{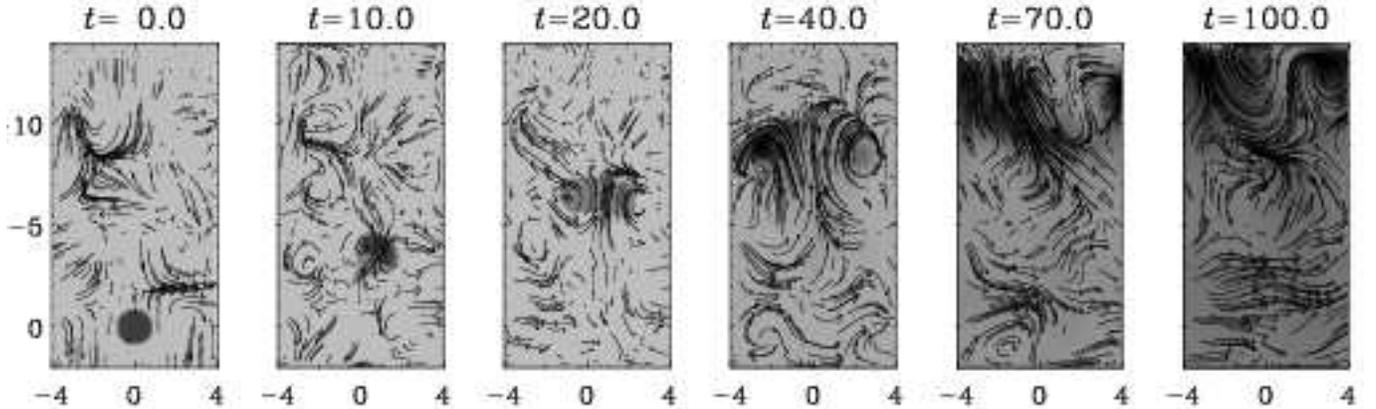}\caption[]{
As for the run shown in \Fig{Fpslice0_n50r1_rad1part_xi30_longer}, but with
strong three-dimensional perturbations. For each time the grey scale
is here adjusted between minimum and maximum values of $s$. Note that
the bubble still makes it all the way to the top of the box, albeit at
a somewhat later time.
}\label{Fpslice0_n50r1_rad1part_xi30_longerb}\end{figure*}

\begin{figure*}[t!]\epsfbox{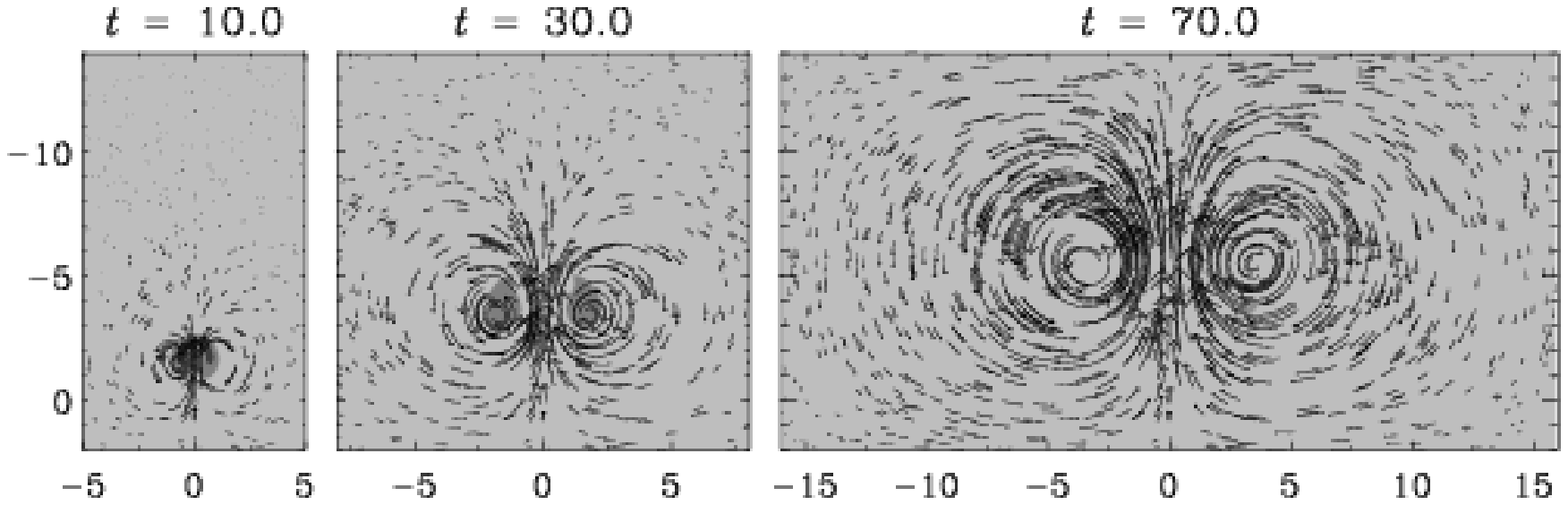}\caption[]{
Velocity vectors superimposed on a grey scale representation of the
entropy for the same case as in \Fig{Fpslice0_n50r1_rad1part_xi30_longer},
but for a two-dimensional cartesian calculation. After $t=20-30$ the
results begin to be affected by boundary effects. This is related
to the strong nonlocality of two-dimensional calculations. Note
that $L_x=16$, but in the first two panels only a smaller range is
shown. $R_0=1.0$. $200\times100$ meshpoints.
}\label{Fpslice0_n200x50r2}\end{figure*}

In \Fig{Fpmass_int_comp} we compare the momentum balance for the two
cases with different values of $z_{\rm top}$. Again, the agreement between
the momentum and the time integrated buoyancy force is good up to $t=10$
(for $z_{\rm top}=-6$) or up to $t=30$ (for $z_{\rm top}=-14$). The presence
of the boundary clearly influences the motion of the bubble; nevertheless the
top of the bubble does manage to reach the boundary, as we see from 
\Fig{Fpslice0_n50r1_rad1part_xi30_longer} (for weak initial perturbations)
and \Fig{Fpslice0_n50r1_rad1part_xi30_longerb} (for strong initial
perturbations). This is in contrast to the
results of Sinjab \ea (1990) who conclude that the blobs
in their calculations `fail to reach the top of the envelope'. In our
calculation the hot part of the bubble seems to stop at some point
just below the boundary and stays there. Furthermore, Sinjab \ea
(1990) interpret their two-dimensional calculations as indicating a
fragmentation. We found no indication of any such effect, irrespective
of whether the stratification was weak or strong. For
comparison we also carry out two-dimensional cartesian calculations verifying
that in the two-dimensional case the eddies do eventually travel downwards
when they come sufficiently close to the horizontal boundaries. Again,
in the weakly stratified case the bubbles can travel to large heights,
but the effects of the boundaries begin to become important much earlier
than in the three-dimensional case; see \Fig{Fpslice0_n200x50r2}.

\epsfxsize=18cm

\epsfxsize=18cm

\epsfxsize=18cm

The results of Sinjab \ea (1990) are of course for the stratified
case. However, our point is that it is the restriction to two dimensions
(in cartesian geometry) that causes boundary effects to become extremely
pronounced. Since stratification itself can also act as an effective
boundary, it was necessary to go to the weakly stratified case where
it is possible to move the boundaries much further away. Although the
box shown in \Fig{Fpslice0_n200x50r2} was big enough to prevent the
edge of the bubble from moving down again, boundary effects did begin to
affect the evolution at times as early as $t=20$.

The ability of our bubble to rise right to the top of the box might
seem surprising when one recalls that atmospheric thermals only manage
to reach a certain maximum height. However one must not forget that
the buoyancy force, relative to gravity, is here orders of magnitude
greater than in the case of a typical atmospheric thermal (see e.g.\
Scorer 1957). A better analogy might therefore be that of an air bubble
rising through water and eventually reaching the surface.

\section{Calculation of the entrainment parameter}

The introduction of a parameter to describe the entrainment of fluid by a rising
`cloud' was proposed by Morton \ea (1956). They introduced an equation of the form
\EQ
\dot{V}=4\pi R^2\alpha_{\rm v} v,
\label{dotV}
\EN
which they regarded as describing `conservation of volume'. We have here
rewritten their Eq.~(16) using a slight change of notation. The `constant'
$\alpha_{\rm v}$ will be referred to hereafter as the volume entrainment
coefficient. The above equation can be simplified to
\EQ
\dot{R}=\alpha_{\rm v}v,
\label{dotR}
\EN
where $R$ is the volume radius.

We note that \Eq{dotV} is taken over in the work of Turner (1963),
with $\alpha_{\rm v}$ becoming Turner's `alpha'.\footnote{We note that
although $\alpha$ has the same meaning in Morton \ea (1956) and
Turner (1963) the characteristic velocities of the bubbles are defined
differently, and this is `compensated' by the inclusion of a $k$-factor
in the entrainment equation of Morton et al.\ ($k=\mbox{ratio}$ of mean
to axial velocity). This would not matter, except that when comparing
with experiment, Morton et al.\ believe the observations relate to the
axial velocity whereas Turner takes them as referring to the mean
velocity. This means that when comparing experimental results it is
really the $\alpha k$ of Morton et al.\ which should be compared
with the $\alpha$ of Turner.}

Now the concept of `conservation of volume' lying behind \Eq{dotV}
will be unfamiliar to many physicists. We therefore thought it to be
worthwhile to concentrate instead on the accretion of mass rather than
of volume and to write
\EQ
\dot{M}=4\pi R^2\bra{\tilde\rho}\alpha_{\rm m} v,
\label{dotM}
\EN
where $\dot{M}$ is the mass entrainment rate, $\bra{\tilde\rho}$ some
average of the surrounding material density near the bubble, and
$\alpha_{\rm m}$ the mass entrainment coefficient. Although an average
of $\tilde\rho$ over the bubble surface might be more appropriate,
it is more convenient (and adequate for present purposes) to use
a volume average. This gives
\EQ
\dot{M}={3\over R}M_{\rm d}\alpha_{\rm m}v
\label{dotM2}
\EN
where, as before, $M_{\rm d}$ is the mass displaced by the bubble.

\begin{figure}[h]\epsfbox{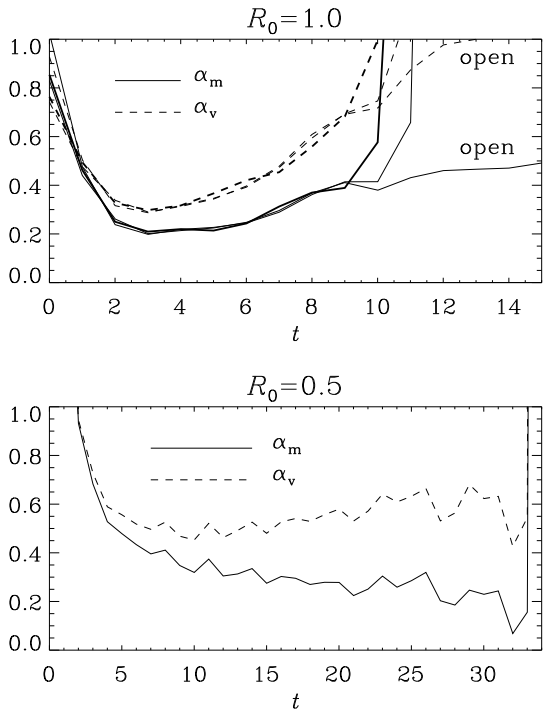}\caption[]{
Entrainment parameters $\alpha_{\rm m}$ (solid line) and $\alpha_{\rm v}$
(dashed line) for the case of strong stratification, $z_{\rm top}=-6.0$
and $\xi_0=0.3$, and two different initial bubble radii. In the upper
panel, thick lines (solid and dashed) refer to the case with $z_{\rm
top}=-6.5$ and $\xi_0=0.1$. Note that in the case of open boundaries
(and $z_{\rm top}=-6.0$; upper panel) the value of $\alpha_{\rm m}$
remains at about 0.4 after $t=10$.
}\label{Fpalpha_comp_strat}\end{figure}

We can now use Eqs~\eq{dotR} and \eq{dotM2} to calculate $\alpha_{\rm
v}$ and $\alpha_{\rm m}$, all other quantities being already known. The
results are shown in \Figs{Fpalpha_comp_strat}{Fpalpha_comp}.

\epsfxsize=8.8cm

\epsfxsize=8.8cm\begin{figure}[h]\epsfbox{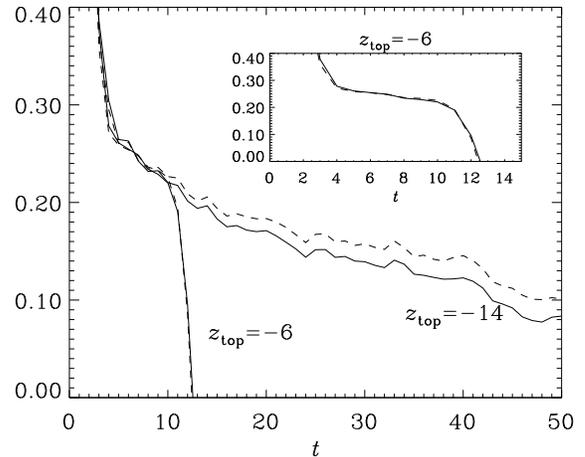}\caption[]{
Entrainment parameters $\alpha_{\rm m}$ (solid line) and $\alpha_{\rm v}$
(dashed line) for the weakly stratified cases with $z_{\rm top}=-14$
($\xi_0=26.8$) and $z_{\rm top}=-6$ ($\xi_0=30$). In the inset one sees
more clearly that for $z_{\rm top}=-6$ there is still some indication of
a plateau in $\alpha_{\rm m}$ and $\alpha_{\rm v}$. $R_0=1.0$.
}\label{Fpalpha_comp}\end{figure}

It is interesting to note that for those parts of the curves not
influenced by the special effects discussed previously, the quantity
$\alpha_{\rm v}$ is in reasonable agreement with the experimental
values of 0.25--0.34 (Morton \ea 1956, Turner 1963), with the agreement
being better in the more strongly stratified case. For smaller initial
bubble radii, $R_0=0.5$, $\alpha_{\rm v}$ attains noticeably larger
values. Nevertheless, for both bubble radii investigated $\alpha_{\rm m}$
is roughly comparable ($\alpha_{\rm m}\approx0.25$). Thus, the assumption
of Turner (1963) that the value of $R_0$ does not influence `alpha'
can be verified only for $\alpha_{\rm m}$.

Better agreement between the `experimental' and `calculated' values cannot
perhaps be expected since the Reynolds numbers for our calculations
are lower than those appropriate to the experimental results, which
leads to a more laminar entrainment process in our case.

Regarding $\alpha_{\rm m}$, we note that the entrainment
of material leads to an effective drag force on the bubble given by
\EQ
\mbox{Drag force}=\dot{M}v=4\pi R^2\tilde\rho\alpha_{\rm m} v^2,
\label{Mdrag}
\EN
which may be compared with the normal hydrodynamic drag
\EQ
\mbox{Hydrodynamic drag}=\half\pi R^2\tilde\rho C_{\rm D} v^2,
\label{Hdrag}
\EN
with $C_{\rm D}$ being the hydrodynamic drag coefficient. According
to Moore (1967), normal hydrodynamic drag plays only a minor r\^ole in
influencing the motion of a thermal. We can now check Moore's point with
the help of \Eqs{Mdrag}{Hdrag}; we confirm that hydrodynamic drag really
does have only a minor influence on the motion, except possibly during
the final stages in the weakly stratified case.

\section{Details of the entrainment}

\begin{figure}[h]\epsfbox{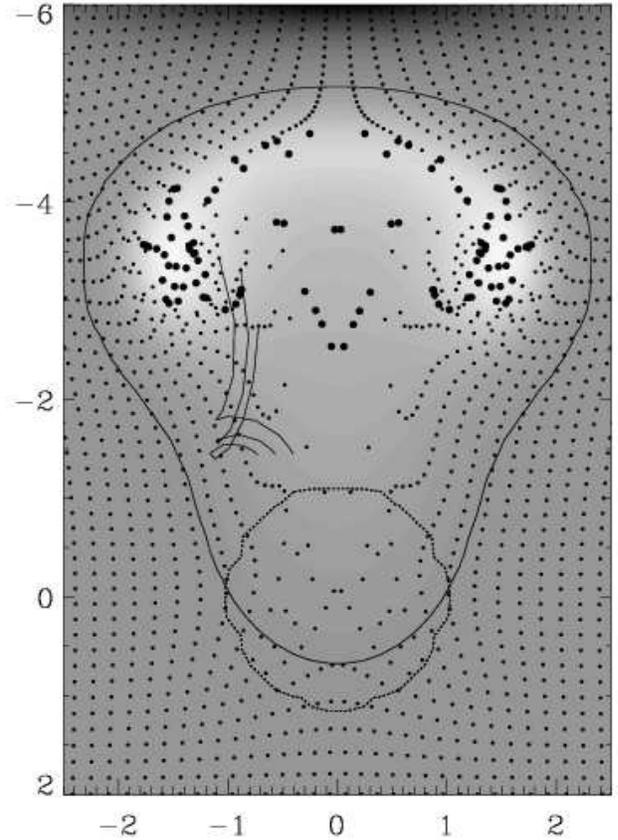}\caption[]{
Tracer particles superimposed on a grey scale representation of entropy
(bright indicates high entropy). Large dots represent particles for which
initially $s\ge s_{\rm crit}$, i.e.\ which originated in the initial
bubble. Small dots represent particles that come from outside the original
bubble. The contours show the position where $s=s_{\rm crit}$ (dashed for
the initial time and solid for $t=10$). The three
neighbouring lines show the particle trajectory of entrained particles
that were originally outside the bubble (at $x=-0.72$, to $-0.40$
and $z=-1.43$, and have now moved upwards to the point indicated by a
small dot). Note that $L_x=4$, but only the range $|x|\leq2.5$ is shown.
$100^2\times50$ meshpoints, $\Delta s=0.5$, $R_0=1.0$.
}\label{Fppart}\end{figure}

Details of the entrainment process are viewed best in terms of tracer
particles that are passively advected by the flow. In \Fig{Fppart}
we show the location of initially uniformly distributed particles at
$t=10$. Particles that were originally inside the bubble (as defined by
$s\ge s_{\rm crit}$) remain inside the bubble for all times. However, an
increasing portion of new particles from outside the bubble is constantly
being entrained, which happens mostly through the top boundary of the
bubble. These particles then move along the periphery of the bubble
tailwards where they find their way into the centre of the ring vortex
associated with the bubble.

\epsfxsize=8.8cm

In \Fig{Fppart} we have also indicated the trajectory of three
neighbouring particles that were originally above the bubble and were
subsequently entrained and lifted upwards together with the bubble. The
kinks in each of these trajectories correspond to the moment at which
the particles were overtaken by the bubble and became entrained.

\section{Conclusions}

The main aim of this paper was to test the validity of the bubble or
thermal concept; in this respect the following conclusions can be drawn.

We found that the bubble could easily be followed as a well-defined
entity throughout the calculations. Nevertheless, its dynamical behaviour
consisted of two distinct phases. In the first of these, the dynamical
behaviour expected from the simple bubble dynamics of equating buoyancy
forces to rate of momentum change was indeed confirmed. However, in the
second phase an unexpected braking effect appeared. Further
investigation led us to attribute this effect to a combination of boundary
effects (artificial) and buoyancy braking (real).

We found, in contrast to the results of Sinjab \ea (1990), that the top
part of the bubble goes on rising until it reaches the surface. Bubbles
penetrating to the surface layers favour the view that dissipation is
an important factor in understanding contact binaries (Hazlehurst 1985,
1996).

Since the code used by us was a three-dimensional one including effects
of viscosity and thermal (radiative) conductivity, we were in a position
to follow in detail and more realistically the mass changes of the
bubble due to entrainment of material. This made it possible for us
to {\it calculate} the value of the entrainment coefficient $\alpha$
rather than, as was previously necessary, to take some value from glider
observations (atmospheric thermals) or experiment. The values found by us
were in good agreement with the those derived empirically, although the
non-constancy of `alpha' means that we have throughout referred to the
entrainment coefficient rather than the entrainment constant. Finally,
the introduction of a mass entrainment coefficient to replace (or at
least supplement) the previously used volume coefficient appears to us
desirable on physical grounds.

\begin{acknowledgements}
We thank Dr Robert Smith for useful comments regarding \Fig{Fppart}.
We also thank an anonymous referee for suggestions which led to an
improvement of the presentation. John Hazlehurst is grateful for the
hospitality of Nordita during this investigation. We acknowledge the
use of GRAND, a high-performance computing facility funded by PPARC.
\end{acknowledgements}

\end{document}